\documentstyle[mprocl]{article}

\begin{document}

\title{MICROLENSING RESULTS AND THE GALACTIC MODELS}

\author{FRANCESCO DE PAOLIS\footnote{Also at the Bartol Research
Institute, Univ. of Delaware, Newark, Delaware, 19716-4793, USA.} 
and PHILIPPE JETZER}
\address{Paul Scherrer Institute, Laboratory for Astrophysics, CH-5232 
Villigen PSI, and
Inst. of Theor. Phys., Univ. of Zurich, Winterthurerstr.
190, CH-8057 Zurich, Switzerland}

\author{GABRIELE INGROSSO}
\address{Dipartimento di Fisica and INFN, Universit\'a di Lecce,
CP 193, I-73100 Lecce, Italy}

\maketitle

\abstracts{
Microlensing results towards the LMC strongly depend on the 
properties of both the luminous and the
dark matter distribution in the Galaxy.
The two main sources of uncertainty come from the poor knowledge of the 
rotation curve at large galactocentric distance and from the possibility 
that MACHOs follow radial orbits in the outer Galaxy.
Given these uncertainties 
MACHO mass values $ \sim 0.1~ M_{\odot}$ are still consistent with 
observations and thus brown dwarfs are to date viable candidates for MACHOs
\cite{modello}. }


\section{Introduction} 
Microlensing events towards the Large Magellanic Cloud (LMC) entail that 
a sizable fraction of the halo dark matter of the Galaxy is in the form of
MACHOs (Massive Astrophysical Compact Halo Objects).
However, although the evidence for MACHOs is firm, robust results are 
only the value of the optical depth $\tau=2.9^{+1.4}_{-0.9} \times
10^{-7}$ reported by the 
MACHO \cite{macho} team and the lower limit $\sim 10^{-2}~M_{\odot}$ to the 
average MACHO mass $m_M$ found by the MACHO and EROS \cite{eros} 
collaborations.
Further results, in particular $m_M$ and the halo mass 
$f_M$ in form of MACHOs, strongly depend on the assumed distributions
of dark and visible matter.
In this respect, usually the standard halo model 
- which assumes spherical symmetry, flat rotation curve and 
Maxwell-Boltzmann statistics - and a ``median'' disk model 
are taken as a baseline for comparison. Within these assumptions 
the MACHO team reported the values $m_M = 0.5^{+0.3}_{-0.2}~M_{\odot}$ 
and $f_M = 50^{+30}_{-20}\%$.

Actually, the previous hypotheses for the distribution 
of dark and visible matter 
can be relaxed and one can consider a set of models 
for the halo - flattened or spherical and/or
in which the MACHOs do not obey the 
Maxwell-Boltzmann statistics - and for the disk - ``maximal'' or ``minimum'' -,
without entering in contradiction 
with the observed galactic rotation curve.

Here we investigate the expected microlensing results 
for a class of halo models 
based on the King-Michie distribution function. 
For these models we determine in a self-consistent 
way the distribution of dark matter 
and study the effect of considering different parameters for the visible part 
of the Galaxy (for more details see \cite{dij}).

\section{Model and microlensing results} 
We assume that the Galaxy contains two main component, namely, the 
visible (stars) and the dark component (MACHOs). 
We consider stars to be distributed according to a central bulge 
and an exponential disk.
Relevant parameters are the bulge mass $M_b = 2\pm 1 \times 10^{10}~M_{\odot}$,
the local projected star mass density $\Sigma_0 = 50 \pm 25~M_{\odot}$ 
pc$^{-2}$ and the disk scale length $h = 3.5 \pm 1$ kpc. 

MACHOs are assumed to be described by the King-Michie distribution function.
As it is well known, this function introduces an energy and angular momentum 
cutoff in the velocity space, which take into account the existence of an 
upper limit to the MACHO velocity at any point in the halo 
(the local escape velocity) and the possibility that MACHOs located in the 
outer part of the halo follow more radial orbits.
Visible and dark components are considered to be in hydrostatic equilibrium 
in the gravitational potential $V$ solution of the Poisson equation,
which we solve assuming spherical symmetry for the dark mass distribution. 
Correspondingly, we obtain the MACHO mass density 
\begin{equation}
\rho_H(r)= A~ (2 \pi \sigma^2)^{-3/2}~ e^{[W(r)-W(0)]}~
\left(\frac{r_a}{r}\right)~
\int_0^{W(r)}[e^{-\xi}-e^{-W(r)}]~F(\lambda)~d\xi ~, 
\label{eq:2.6}
\end{equation}
where $W(r)=- V(r)/\sigma^2$ is the energy cutoff parameter,
$r_a$ the anisotropy radius,
$\lambda=(r/r_a)\sqrt{\xi}$ and $F(\lambda)$ is the Dawson integral.
$A$ is a normalization constant and $\sigma$ is the one-dimensional
MACHO velocity dispersion.

The rate at which a single star is microlensed is given by 
\cite{djm,griest}
\begin{equation}
\Gamma=
2 D r_E \frac{\rho_0}{M_{\odot}} \frac{1}{\sqrt{\bar \mu}}
\int_0^{+v_c} dv_T \int^1_0 d\tilde x~ v_T^2 f(\tilde x,v_T) 
[\tilde x(1-\tilde x)]^{1/2} H(\tilde x)~,
\label{eq:ta}
\end{equation}
where all MACHOs have been 
assumed of the same mass $\bar\mu$ (in solar units), 
$D$ is the LMC distance, $r_E= \sqrt{4GM_{\odot}D}/c$,
$H(\tilde x)= \rho_H(\tilde x)/\rho_0$ ($\rho_0$ is the local mass density)
and  
$f(\tilde x,v_T)$ is the projection of the MACHO velocity distribution 
function in the plane perpendicular to the line 
of sight. For an experiment monitoring $N_{\star}$ stars during an
observation time $t_{obs}$ the total number of expected events will be
$N_{ev}=N_{\star} t_{obs} \Gamma$. 
The expected event duration $T$ is defined as \cite{djm} 
\begin{equation}
T=\frac{r_E} {v_0} \sqrt{\bar\mu} \frac{\gamma(1)}{\gamma(1/2)}~,
\end{equation}
where the $\gamma(m)$ functions are given by
\begin{equation}
\gamma(m) \equiv \int_0^{+v_c} dv_T
\int_0^1 d\tilde x \left(\frac{v_T} {v_0}\right)^{2-2m} v_T
f(\tilde x,v_T)  [\tilde x(1-\tilde x)]^m H(\tilde x)~.
\label{eq:wf}
\end{equation}
Finally, the average MACHO mass $\bar M$ is given by
\begin{equation}
\bar M = \frac{<\tau^1>}{<\tau^{-1}>} \frac{\gamma(0)}{\gamma(1)}~,
\label{eq:mt}
\end{equation}
where $<\tau^1>$ and $<\tau^{-1}>$ are determined through the
observed microlensing events

Our model results are given in Table 1,
where the dark matter parameters are the core radius $a$, 
the local mass density $\rho_0$ and the anisotropy radius $r_a$.
We consider 
three models of luminous matter corresponding to the ``minimum'' 
(rows 1 and 4), the ``median'' (rows 2 and 5) and the ``maximum'' 
(rows 3 and 6) disk model. Here we take only models which lead to 
a flat rotation curve in the region $5~{\rm kpc}<r < 50~{\rm kpc}$.
\begin{table}
\caption{Mean values of parameters and microlensing results for
models with flat rotation curve up to the LMC.
We take an average MACHO mass of $0.1 M_{\odot}$,
an observation time of 1 year and a number of $10^6$ monitored stars.}
\vskip 0.3cm
\begin{tabular}{|c|c|c|c|c|c|c|}
\hline
$a$&$\rho_0/10^{-3}$&$r_a$&$M_H^{~LMC}$&
$N_{ev}$ & $T$ & $[\gamma(0)/\gamma(1)]$ \\
(kpc) & $(M_\odot~{\rm pc}^{-3})$ & (kpc)  
& $(10^{11}~M_\odot)$ & &
(days) &  \\
\hline
$4.9 \pm 0.9$ & $9.1 \pm 0.9$ & $\infty$ & $6.0 \pm 0.9$ & $6.4\pm 0.9$ &$ 22.6\pm 1.0$ & $8.2 \pm 0.8$ \\
$6.6 \pm 1.9$ & $6.4 \pm 0.7$ & $\infty$ & $6.3 \pm 0.9$ & $5.9\pm 0.9$ &$ 21.7\pm 1.0$ & $8.5 \pm 1.0$ \\
$13  \pm 4  $ & $4.0 \pm 0.5$ & $\infty$ & $6.9 \pm 1.0$ & $5.8\pm 0.9$ &$ 19.2\pm 1.0$ & $10.2 \pm 1.1$ \\
\hline 
$5.7\pm 1.5$ & $9.7 \pm 1.1$ & $40\pm 15$ & $4.9 \pm 0.7$ &$5.7\pm 0.8$ & $25.1 \pm 0.9$ & $6.9 \pm 0.6$ \\
$8.9\pm 3.5$ & $7.0 \pm 1.0$ & $42\pm 16$ & $5.1 \pm 0.8$ &$5.2\pm 0.8$ & $24.1 \pm 1.3$ & $7.3 \pm 1.1$ \\
$17 \pm 3  $ & $4.5 \pm 0.6$ & $45\pm 13$ & $5.8 \pm 0.7$ &$5.0\pm 0.9$ & $21.6 \pm 1.0$ & $8.6 \pm 0.8$ \\
\hline
\end{tabular}
\end{table}

Results for the standard halo model 
are given in the lines 1-3. The radial anisotropy (lines 4-6)  
has the effect to decrease $N_{ev}$
and $[\gamma(0)/\gamma(1)]$ (which is related to the expected 
average MACHO mass), while the event duration $T$ gets increased.
The decrease of $N_{ev}$ is mainly due to the reduction 
of the halo mass  $M_H^{~LMC}$ up to the LMC distance, 
while the reduction of $v_T$ leads to an increase of $T$.

Till now, we assumed a flat rotation curve in the range $5-50$ kpc.
Actually, the galactic rotation curve is well measured only in the range 
$5-20$ kpc.
Thus, to select acceptable physical models one should require that \cite{gates}:
i) the total variation in $v_{rot}(r)$ in the range $5~{\rm kpc}<r<20$ kpc 
is less than 14\%; 
ii) the rotational velocity $v_{rot}(LMC)$ at the LMC 
is in the range  $150-307$ km s$^{-1}$.
With these assumptions we find that microlensing results strongly 
depend on $v_{rot}(LMC)$ and can vary up to a factor $\sim 6$
(see Figs. 1 and 3 in \cite{dij}).

In conclusion the variation in the expected microlensing results is at least 
within 30\% from the value one gets for the standard halo model, due to anisotropy 
effect. This factor strongly increases if one allows for less 
restrictive conditions on the galactic rotation curve.

\end{document}